\documentclass[amsmath,pra,twocolumn]{revtex4}
\usepackage{times}
\usepackage{epsfig}
\usepackage{amssymb}
\usepackage{amsmath}
\usepackage{amsfonts}
\usepackage{bm}

\begin{document}
	
	\title{Natural line profile asymmetry}
	\author{A. Anikin$^{1,2}$, T. Zalialiutdinov$^{1,3}$, D. Solovyev$^{1}$  }
	
	\affiliation{ 
		$^1$ Department of Physics, St.Petersburg State University, St.Petersburg, 198504, Russia
		\\
		$^2$ D. I. Mendeleev Institute for Metrology, St. Petersburg, 190005, Russia
		\\
		$^3$ National Research Center, Kurchatov Institute, B.P. Konstantinov Petersburg Nuclear Physics Institute, Gatchina, Leningrad District 188300, Russia}
	
	\begin{abstract}
		The paper discusses the line profile asymmetry of the photon scattering process that arises naturally in quantum electrodynamics (QED). Based on precision spectroscopic experiments conducted on hydrogen atoms, we focus our attention on the two-photon $1s-2s$ transition. As one of the most precisely determined transition frequencies, it is a key pillar of optical frequency standards and is used in determining fundamental physical constants, testing physical principles, and searching constraints on new fundamental interactions. The results obtained in this work show the need to take into account the natural line profile asymmetry in precision spectroscopic experiments.
	\end{abstract}
	
	\maketitle
	\section{Introduction}
	
	Precise measurements of transition frequencies in various atomic systems are of paramount importance in modern physics. By providing a detailed interpretation of a theory, spectroscopic experiments are able to identify effects that can be used to enhance our understanding of physical processes, test theoretical hypotheses, or verify fundamental physical principles. For such attractive astrophysical applications as the search for Dark Matter \cite{Kennedy} or the variation in time of physical constants \cite{DIRAC1937,Webb}, the laboratory measurements of transition frequencies with increasing accuracy are required. Another important aspect of improving the accuracy of the experiment is the determination of the fundamental physical constants and the definition of frequency standards. The leaders for the latter are atomic clocks, which have an unprecedented relative error of $2.1\times 10^{-18}$, see, e.g., \cite{AtCl-Cs,AtCl-Sr}. Laboratory experiments with simple atomic systems such as hydrogen or helium atoms have also demonstrated fast progress \cite{Parthey,Mat,van2011frequency}, where relative uncertainties were reported at the order of $4.5\times 10^{-15}$ and $8\times 10^{-12}$, respectively.
	
	Requiring increasingly subtle effects to be taken into account, experiments with hydrogen have reached a level where the spectral line profile has become an observable quantity \cite{H-exp}. Theoretical derivation of the line shape can be obtained within the framework of quantum mechanics (QM) or more rigorously within the framework of QED (quantum electrodynamics) theory. According to F. Low \cite{Low}, the Lorentz line profile in QED arises in a natural way within the resonance approximation. Until recently, the resonant approximation was sufficient for the above purposes, although theoretical studies were also carried out for effects lying beyond its limits, see \cite{Andr,ZSLP-report,AZSL-2022} and references therein. 
	
	In particular, it was shown that the nonresonant contributions arising from the differential or total cross section of the photon scattering process lead to a significant line profile distortion. This, in turn, affects the determination of the transition frequency. Being close to the experimental scheme, the quantum interference effect (QIE), responsible for nonresonant corrections to the transition frequency, has been widely studied in applications to various atomic systems for one-photon scattering \cite{Jent-Mohr,HH-2010,HH-2011,Sansonetti,Brown-2013,MHH-2015,Amaro-2015,Amaro-mH-2015} and recently for two-photon absorption experiments \cite{AZSL-2022, Anikin,Anikin2021}.
	
	The main feature of QIE is dependence on the angles between the directions (or polarizations) of the absorbed and emitted photons involved in the process. This opens up the possibility of avoiding such effects by choosing the geometry of the experimental setup or by using the asymmetry parameter in the appropriate fitting of the line profile fit \cite{H-exp}. While remaining significant, nonresonant corrections cannot be always reduced to zero by the choice of geometry: for example, corrections arising through the total cross section \cite{LSPS,PRA-LSPS}, for specific states of atoms in the differential cross section \cite{Anikin} or for a special experimental scheme \cite{Schwob,deB-0,deB-1,deB-2,Anikin2021}. On this basis, it can be argued that the procedure of fitting the observed line shape by the profile, having regard to the asymmetry parameter, remains the immutable way to determine the transition frequency precisely. In this case, it becomes necessary to calculate the corresponding asymmetry parameter (nonresonant correction) for each particular experiment.
	
	In this paper, we will focus on another type of line profile asymmetry. In contrast to QIE and other nonresonant corrections, where resonant and nearest neighboring states can be resolved experimentally, this asymmetry is the result of deriving the line profile beyond the resonant approximation and naturally arises for specific transitions, i.e., one-photon electric, magnetic dipole or higher multipole transition rates (with the same concepts for multiphoton processes).
	
	\section{Natural line profile asymmetry}
	\label{na}
	
	The natural asymmetry of the line profile was originally discussed in \cite{PRA-LSPS}, where it was shown that the level width does not depend on frequency only within the resonant approximation. As an example, the profile of the Lyman$_\alpha$ line in the hydrogen atom was considered in \cite{PRA-LSPS} (see also \cite{Andr}), where the "velocity" form of the level width was found as $\Gamma(\omega)=(2^{11}/3^9)\alpha^3\omega$ (in atomic units, a.u., $\alpha$ is the fine structure constant) instead of the resonant one $\Gamma(\omega)=(2/3)^8\alpha^3$ a.u. Alongside this, the former initially arises in the QED theory, while the latter is the result of the approximation, see \cite{AZSL-2022} for details and further extensions. This immediately leads to (by inserting $\Gamma(\omega)$) a modification of the Lorentz profile, $L(\omega)$, which becomes asymmetric. An investigation of the frequency dependence of the numerator of the Lorentzian line profile can also be found in \cite{PhysRevA.25.3079}, along with consideration of cascade emission/absorption processes. 
	
	The asymmetry parameter can be considered as an additional frequency shift in the approximation of its smallness with respect to the level width. To determine the correction to the transition frequency, it suffices to fulfill the extremum condition $dL(\omega)/d\omega=0$, from which the frequency can be found as $\omega=\omega_0+\delta_a$. Here $\omega_0$ represents the resonant quantity, and $\delta_a$ is responsible for the asymmetry (due to the different frequency dependence of the line profile). The additional asymmetry shift $\delta_a$, which follows from the "velocity" form of $\Gamma(\omega)$, reads
	\begin{eqnarray}
		\label{1}
		\delta_a = \frac{1}{8}\Gamma(\omega_0)\left[\frac{d\Gamma(\omega)}{d\omega}\right]_{\omega=\omega_0}.
	\end{eqnarray}
	For the Lyman$_\alpha$ line in the hydrogen atom it gives $1$ Hz with the parametric estimation of $(\alpha Z)^6$ a.u.
	
	The value of $\delta_a$ lies beyond the current accuracy of the Lyman$_\alpha$ frequency, which is at the level of few MHz \cite{Eikema}. The correction (\ref{1}) can easily be extended to the case of two-photon $1s-2s$ absorption \cite{Parthey,Mat}, where the relative experimental error is of the order $10^{-15}$. For this, the estimates $\Gamma_{2s}^{(2\gamma)}(\omega_0)\sim m\alpha^2(\alpha Z)^6$ and $\omega_0\sim m(\alpha Z)^2$ in relativistic units (r.u.) lead to a negligible value of the order of $\left[m\alpha^2(\alpha Z)^6\right]^2/m(\alpha Z)^2 = m\alpha^4(\alpha Z)^{10}$ r.u. or $\alpha^2(\alpha Z)^{10}$ a.u. However, considering the explicit frequency dependence of the level width for the two-photon transition rate \cite{LabKlim,Zon}, see also \cite{Andr}, we find the following expression
	\begin{eqnarray}
		\label{2-1}
		\Gamma^{(2\gamma)}_{2s}(\omega, \omega') \sim \omega^3 \omega^{\prime 3} \times 
		\\
		\nonumber
		\times \int d\boldsymbol{n}_{\boldsymbol{k}}d\boldsymbol{n}_{\boldsymbol{k}^{\prime}} \Bigg{|} \sum_{\boldsymbol{e}\boldsymbol{e}^{\prime}} \sum_n \Bigg{(} \frac{\langle 2s |A_{\mathrm{abs}}| n \rangle\langle n |A^{\prime}_{\mathrm{abs}}| 1s \rangle}{E_n - E_{1s} + \omega} +
		\\
		\nonumber
		+ \frac{\langle 2s |A^{\prime}_{\mathrm{abs}}| n \rangle\langle n |A_{\mathrm{abs}}| 1s \rangle}{E_n - E_{1s} + \omega^{\prime}} \Bigg{)}\Bigg{|}^2,
	\end{eqnarray}
	in which $A_{\rm abs} = (\boldsymbol{e}\boldsymbol{\alpha})e^{-\mathrm{i}\boldsymbol{k}\boldsymbol{r}}$ and $A^{\prime}_{\rm abs} = (\boldsymbol{e}^{\prime}\boldsymbol{\alpha})e^{-\mathrm{i}\boldsymbol{k}^{\prime}\boldsymbol{r}}$ denotes the wave functions of the absorbed photons, where $\boldsymbol{e},\boldsymbol{e}^{\prime}$ -- polarization vectors, $\boldsymbol{k},\boldsymbol{k}^{\prime}$ -- wave vectors ($|\boldsymbol{k}|\equiv \omega$ and $|\boldsymbol{k}^{\prime}|\equiv \omega^{\prime}$ -- photon frequencies) and $\boldsymbol{\alpha}$ represents the corresponding Dirac matrix. Summation in (\ref{2-1}) runs over photons' polarization vectors and over entire spectrum of Dirac equation. In this expression one can neglect the $\omega$-dependence in the energy denominators by putting $\omega = \frac{\omega_0}{2}$, so that $\Gamma^{(2\gamma)}_{2s}(\omega)$ can be written in the form
	\begin{eqnarray}
		\label{2}
		\Gamma^{(2\gamma)}_{2s}(\omega) = \left(\frac{\omega}{\omega_0/2}\right)^6\Gamma^{(2\gamma)}_{2s}\left(\frac{\omega_0}{2}\right).
	\end{eqnarray}
	where we took into account that the excitation process occurs for two equivalent photons ($\omega = \omega^{\prime}$) at the half-resonant frequency \cite{Parthey,Mat}.
	
	Then, the asymmetry shift arises from the extremum condition for the profile
	\begin{eqnarray}
		\label{3}
		\frac{dL(\delta_a)}{d\delta_a} &=& \frac{d}{d\delta_a} \frac{\left(\frac{\delta_a+\omega_0}{\omega_0}\right)^6\Gamma^{(2\gamma)}_{2s}\left(\frac{\omega_0}{2}\right)}{\delta_a^2 + \frac{1}{4}\left(\left(\frac{\delta_a+\omega_0}{\omega_0}\right)^6  \Gamma^{(2\gamma)}_{2s}\left(\frac{\omega_0}{2}\right)\right)^2} =0,
		\\
		\label{4}
		\delta_a &\approx& -\frac{3\left(\Gamma^{(2\gamma)}_{2s}\left(\frac{\omega_0}{2}\right)\right)^2}{4\omega_0}.
	\end{eqnarray}
	Here $\delta_a \equiv 2\omega-\omega_0$ can be evaluated using the value of $\Gamma^{(2\gamma)}_{2s}\approx 8 s^{-1} \approx 1 $ Hz (see, for example, \cite{Jentschura_2007}) and $\omega_0 \approx 10^{16} $ Hz \cite{Parthey,Mat} which gives a negligible value of about $10^{-14}$ Hz. Nevertheless, this result is more of a model character, since the general process of measuring the $1s-2s$ transition frequency requires a more detailed theoretical description. We will address it in the next section. It should be noted that similar approximation for the asymmetry shift of the $2s-ns/nd$ (here $n=3,\,4,\,6,\,8,\,12$ is the principal quantum number) transition frequencies are valid \cite{Schwob,deB-0,deB-1,deB-2,GM-1s-3s,Mat-2s-8s}.
	
	
	
	The smallness of the natural asymmetry shift arises from the width of the $2s$ metastable state level in hydrogen. However, for the $ns/nd$ states the level width is formed mainly by the one-photon decay rates to the lower states, $\Gamma_{ns/nd}=\sum\limits_{k<n}W^{(1\gamma)}_{ns/nd\rightarrow kp}$. The frequency dependence in "velocity" form can be introduced through the energy conservation law for the scattering process: $E_f+\omega_{\rm em}-2\omega-E_i=0$, where $E_i$, $E_f$ are the energies of the initial and final states (in our case $E_i=1s$ or $2s$ and $E_f=kp$), respectively, and $\omega_{\rm em}$ is the frequency of the emitted photon. Then substitution of the resonance energy $\pm E_{ns/nd}$ yields $\omega_{\rm em} = E_{ns/nd}-E_{kp}+\delta_a$, resulting in the line profile
	\begin{eqnarray}
		\label{5}
		L(\delta_a) = \frac{\left(\frac{\delta_a+\omega_0}{\omega_0}\right)^6\Gamma^{(2\gamma)}_{ns/nd}\left(\frac{\omega_0}{2}\right)}{\delta_a^2 + \frac{1}{4}\left[\sum_{k<n}\left(\frac{\Delta E_{ns/nd,kp}+\delta_a}{\Delta E_{ns/nd,kp}}\right) W^{(1\gamma)}_{ns/nd\rightarrow kp}\right]^2}.
	\end{eqnarray}
	
	Employing the extremum condition, the natural asymmetry shift can be found as
	\begin{eqnarray}
		\label{6}
		\delta_a \approx \frac{3}{4}\frac{\Gamma_{ns/nd}^2}{\omega_0}-\frac{\Gamma_{ns/nd}}{4}\sum\limits_{k<n}\frac{W^{(1\gamma)}_{ns/nd\rightarrow kp}}{\Delta E_{ns/nd,kp}}.
	\end{eqnarray}
	Numerical values of the correction (\ref{6}) are collected in Table~\ref{tab:1}.

	As follows from Table~\ref{tab:1}, the natural line asymmetry is much less than the actual measurement accuracy, which for all considered excited states is of the order of several kHz. The maximum value of about $10^{-2}$ Hz can be found for the $3d$ and $4d$ states. The main conclusion, however, is that the correction depends on the level width, which, in turn, depends on the experimental conditions. In the next section, utilizing the example of $1s-2s$ two-photon absorption, we will demonstrate the importance of a detailed description of the experimental setup for analyzing the asymmetry of the natural line profile.

	\section{$1s-2s$ absorption profile asymmetry}
	\label{2ph}
	
	According to the experiment on measuring the $1s-2s$ transition frequency \cite{Parthey,Mat}, the de-excitation process of the metastable $2s$ state occurs due to the admixture of the $2p$ state in an external electric field. Thus, the level width of the $2s$ atomic state is determined by the decay rate in the presence of mixed state, which can be found in \cite{Ans} (see also \cite{Solovyev_2010,Sol-2015}). It is known that the electric field strength of $475$ V/cm completely mixes the $2s$ and $2p$ states, which leads to the decay of the $2s$ state at a rate equal to the $2p-1s$ transition probability \cite{Ans}. Then, an estimation for the natural asymmetry shift can be obtained from Eq. (\ref{4}) by substituting the magnitude of $\Gamma^{(1\gamma)}_{2p}$ instead of $\Gamma^{(2\gamma)}_{2s}$, result is $-3$ Hz. 
	
\begin{figure}[hbtp]
		\centering
		\caption{The Feynman graph corresponding to the schematic description of the photon scattering process $1s+2\gamma\rightarrow \overline{2s}\rightarrow 1s+1\gamma$ in an external electric field ($\overline{2s}$ denotes the mixed $2s$, $2p$ state). Wavy lines with arrows represent emitted ($\omega_{\rm em}$) or absorbed ($\omega$) photons. The dashed line represents the interaction with an external field. The double solid line denotes the bound electron (Furry picture), the final and initial states denoted as $f$ and $i$, respectively, correspond to the $1s$ state in hydrogen, the intermediate resonance states are given by the $2s$ and $2p$ atomic levels.
		}
		\includegraphics[width=0.8\linewidth]{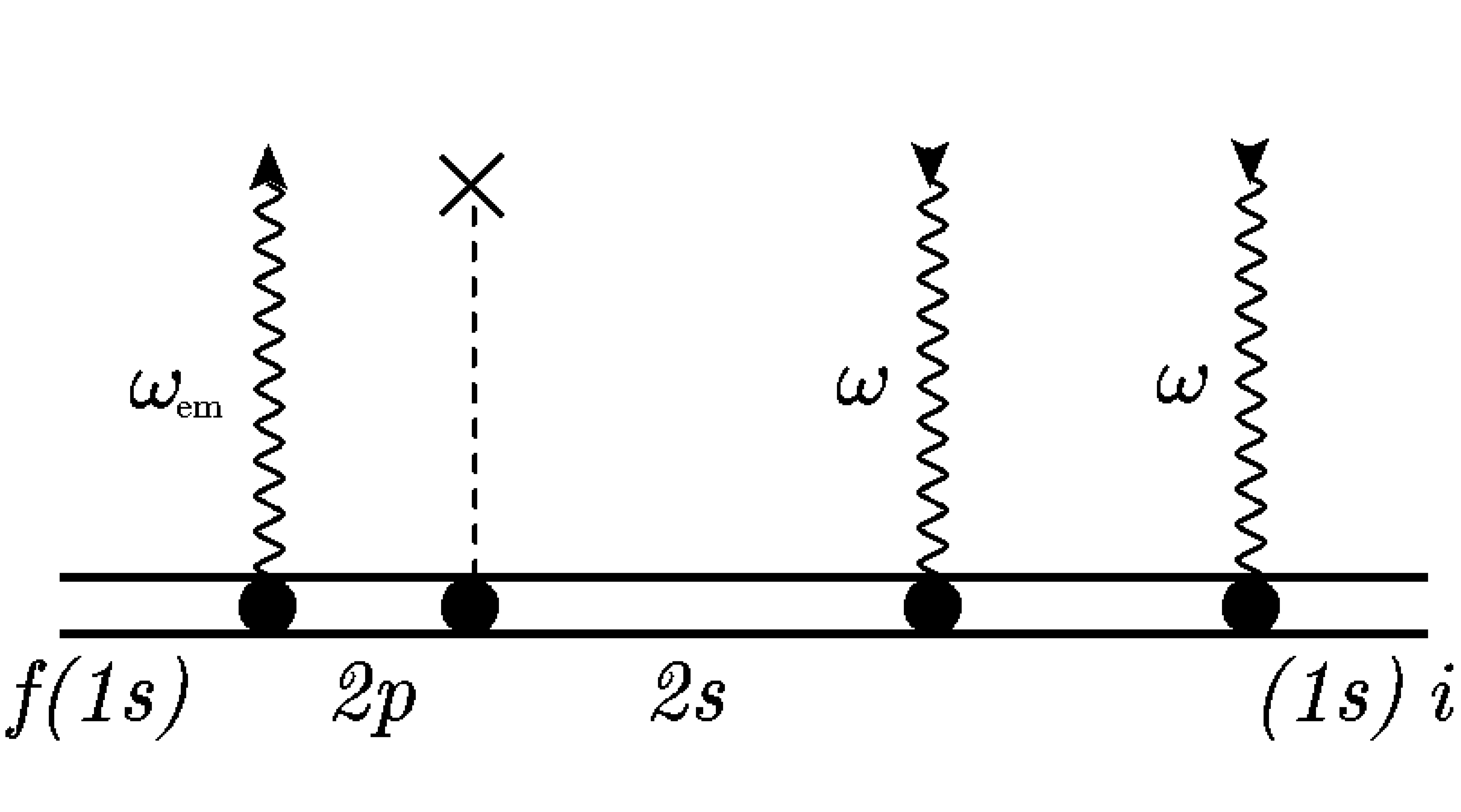}
		\label{fig1}
	\end{figure}

	
	This estimation is of the order of the experimental error of $10$ Hz. However, the width of the $2s$ mixed state should depend quadratically on the electric field strength \cite{Ans,Solovyev_2010,Sol-2015}. For example, decreasing the field strength to $10$ V/cm used in the experiment \cite{Mat} immediately gives a value of $-6\times10^{-7}$ Hz. Varying strongly with the field strength these estimations demonstrate the necessity of an accurate analysis. To do this, we use the $S$-matrix formalism for the Feynman diagram shown in Fig.~\ref{fig1}.
	
	The time which takes atoms to propagate between the excitation and the deexcitation regions, presented in experiments like \cite{Parthey,Mat}, deserves special discussion. Since the $2s$-state for which two-photon absorption occurs is metastable, it makes it possible to separate the excitation region from the interaction with the external electric field. But due to the fact that the action of an electric field leads to immediate decay of the excited $2s-$state and that propagation time, which lies in the region $\tau\in [10,1610]$ $\mu$s \cite{Mat}, is much smaller then the $2s-$state lifetime,  during the propagation between these two regions one should estimate the level width as the inverse to the propagation time. However, the estimation of the level width in an electric field should rather be attributed to the calculation of the transition probabilities \cite{Ans,Solovyev_2010,Sol-2015}, and the delay corresponds to the control of the $2s$ population in time. The latter can be considered according to the exponential decay law with the natural level width of the state $2s$ (assuming that during this time nothing happens to the atom). Thus, the additional factor $\exp(-\Gamma_{2s}^{(2\gamma)}\tau)$ ranges from 1 to $0.987$ and is therefore irrelevant for our estimates.
	
	
	Performing a cumbersome but standard calculation for the $S$-matrix element corresponding to Fig.~\ref{fig1}, we find
	\begin{eqnarray}
		\label{7}
		S_{fi}=-2\pi \mathrm{i}\delta(E_f+\omega_{\rm em}-2\omega-E_i)\frac{e^3(2\pi)^{3/2}}{\omega\sqrt{\omega_{\rm em}}}\qquad
		\\
		\nonumber
		\sum\limits_{n_1 n_2 n_3}\frac{\langle f | A^*_{\rm em} |n_1\rangle\langle n_1| (e\boldsymbol{D}\boldsymbol{r}) |n_2\rangle}{[E_f+\omega_{\rm em}-E_{n_1}+\mathrm{i}0][E_f+\omega_{\rm em}-E_{n_2}+\mathrm{i}0]}
		\\
		\nonumber
		\times\frac{\langle n_2| A_{\rm abs}  |n_3\rangle\langle n_3 | A_{\rm abs}  | i\rangle}{E_i+\omega-E_{n_3}+\mathrm{i}0}.\qquad
	\end{eqnarray}
	Here $A_{\rm em} = (\boldsymbol{e}^*\boldsymbol{\alpha})e^{-\mathrm{i}\boldsymbol{k}_{\rm em}\boldsymbol{r}}$, $A_{\rm abs} = (\boldsymbol{e}\boldsymbol{\alpha})e^{-\mathrm{i}\boldsymbol{k}\boldsymbol{r}}$ denotes the wave functions of the emitted and absorbed photons, respectively, where $\boldsymbol{e}$ -- polarization vector, $\boldsymbol{k}$ -- wave vector ($|\boldsymbol{k}|\equiv \omega$ -- photon frequency) and $\boldsymbol{\alpha}$ represents the corresponding Dirac matrix, $\boldsymbol{D}$ denotes the electric field strength, and the scalar product $(e\boldsymbol{D}\boldsymbol{r})$ represents the dipole interaction of the bound electron with an external field. We also used that the excitation process occurs by two equivalent photons with the same frequencies $\omega$. Summation over $n_1,n_2,n_3$ runs over entire spectrum of Dirac equation including the continuum.
	
	For an accurate description, it is necessary to consider all variants of the interaction of an atom with photons, i.e. to take into account permutations of photon lines in Fig.~\ref{fig1}. Proceeding in this way, it can be found that 1) the two-photon absorption amplitude can be arranged with the last energy denominator in Eq. (\ref{7}) in conjunction with the summation over $n_3$, and 2) all other terms representing the nonresonant contribution, see \cite{Andr,ZSLP-report,AZSL-2022}, can be omitted for our purposes. To form the line profile, we fix the states $n_2=2s$ and $n_1=2p$ (resonant approximation), which yields
	\begin{eqnarray}
		\label{8}
		U_{fi} = 
		\frac{\frac{e^3(2\pi)^{3/2}}{\omega\sqrt{\omega_{\rm em}}}\langle f| A^*_{\rm em}| 2p\rangle\langle 2p|(e\boldsymbol{D}\boldsymbol{r})|2s\rangle A^{(2\gamma)}_{2s,1s}}{[E_f+\omega_{\rm em}-E_{2p}+\mathrm{i}0][E_f+\omega_{\rm em}-E_{2s}+\mathrm{i}0]}.
	\end{eqnarray}
	Here the amplitude $U_{fi}$ is defined via $S_{fi} = -2\pi\mathrm{i}\delta(E_f+\omega_{\rm em}-2\omega-E_i)U_{fi}$ and $A^{(2\gamma)}_{2s,1s}$ denotes the two-photon absorption amplitude \cite{LabKlim}, summation over projections of intermediate states is assumed.
	
	Using the energy conservation law, we can find that the energy denominators in Eq. (\ref{8}) diverge. The regularization of these divergences can be performed in the framework of the QED theory by inserting one-loop self-energy corrections in Fig.~\ref{fig1} \cite{Low,Andr}. In contrast to the case with no external fields the self-energy operator for the denominator with $E_{2s}$ should be averaged over the mixed state $\overline{2s}$ \cite{Ans}, while for the denominator with $E_{2p}$, it should be averaged over the pure $2p$ state. Then, the imaginary part of the averaged one-loop self-energy operator leads to the corresponding level widths. Finally, the amplitude $U_{fi}$ takes the form
	\begin{eqnarray}
		\label{9}
		\frac{\frac{e^3(2\pi)^{3/2}}{\omega\sqrt{\omega_{\rm em}}}\langle f| A^*_{\rm em}| 2p\rangle\langle 2p|(e\boldsymbol{D}\boldsymbol{r})|2s\rangle A^{(2\gamma)}_{2s,1s}}{\left[E_f+\omega_{\rm em}-E_{2p}-\frac{\mathrm{i}\Gamma_{2p}}{2}\right]\left[E_f+\omega_{\rm em}-E_{2s}-\frac{\mathrm{i}\Gamma_{\overline{2s}}}{2}\right]}.
	\end{eqnarray}
	
	By squaring the modulus of the expression (\ref{9}), one can find the line shape of considered process. It should be noted especially that substituting $\omega_{\rm em}=E_{2s}-E_f$ into the first energy denominator of Eq. (\ref{9}) leads directly to the result used in \cite{Ans}, i.e. to the representation of the mixed state in the form $|\overline{2s}\rangle = |2s\rangle+\eta\langle 2p| (e\boldsymbol{D}\boldsymbol{r})|2s\rangle|2p\rangle$, with $\eta = (\Delta E_L-\mathrm{i}\Gamma_{2p}/2)^{-1}$, see also \cite{Solovyev_2010,Sol-2015}. In turn, the energy conservation law given by the $\delta$-function in Eq. (\ref{7}), $E_f+\omega_{\rm em}=2\omega+E_i$, in conjunction with $\omega_{\rm em}=E_{2s}-E_f$ results in the Lorentz contour when the resonant approximation is employed for numerator. In other words, 
	\begin{eqnarray}
		\label{10}
		L(\omega) \sim \frac{W_{2p,1s}^{(1\gamma)}\left|\langle 2p|(e\boldsymbol{D}\boldsymbol{r})|2s\rangle\right|^2 W_{2s,1s}^{(2\gamma)}}{[\Delta E_L^2+\frac{1}{4}\Gamma_{2p}^2][(2\omega-E_{2s}+E_{1s})^2+\frac{1}{4}\Gamma_{\overline{2s}}^2]},
	\end{eqnarray}
	where $\Delta E_L=E_{2s}-E_{2p}$ is the Lamb shift, $W_{2p,1s}^{(1\gamma)}$ is the one-photon $2p\rightarrow 1s$ transition probability and $W_{2s,1s}^{(2\gamma)}$ corresponds to the two-photon absorption rate $1s\rightarrow 2s$.
	
	It is easy to see that the line profile (\ref{10}) is symmetrical and gives maximum at $\omega=(E_{2s}-E_{1s})/2$. The same result can be obtained for $\omega_{\rm em}$ considering that for the case $E_f=E_i=1s$ $\omega_{\rm em}=2\omega$. However, according to (\ref{9}), in order to take into account the natural line profile asymmetry, the frequency dependence should be preserved in both energy denominators and in the numerator. By discarding all unimportant constants, we can write
	\begin{eqnarray}
		\label{11}
		L(\delta_a)\sim \frac{(\delta_a+\omega_0)^7}{\left[\omega_0^2\, \delta_a^2+\frac{1}{4}\Gamma_{\overline{2s}}^2\omega_{\rm em}^2\right]\left[\omega_0^2(\delta_a+\Delta E_L)^2+\frac{1}{4}\Gamma_{2p}^2\omega_{\rm em}^2\right]},
	\end{eqnarray}
	where $\omega_0=E_{2s}-E_{1s}$ and $\delta_a\equiv \omega_{\rm em}-\omega_0$ or $\omega_{\rm em} = \delta_a+\omega_0$.
	
	Using the extremum condition with respect to $\delta_a$ for the profile (\ref{11}), one finds the natural asymmetry shift to be
	\begin{eqnarray}
		\label{12}
		\delta_a \approx \frac{5}{8}\frac{\Gamma_{\overline{2s}}^2}{\omega_0} - \frac{1}{4}\frac{\Gamma_{\overline{2s}}^2}{\Delta E_L} + \frac{1}{16}\frac{\Gamma_{\overline{2s}}^2\Gamma_{2p}^2}{\Delta E_L^3} - \frac{1}{16}\frac{\Gamma_{\overline{2s}}^2\Gamma_{2p}^2}{\Delta E_L^2 \omega_0}.
	\end{eqnarray}
	The difference in the coefficients for the first term in Eq. (\ref{12}) and Eq. (\ref{4}) is explained by the frequency dependence of the numerators and denominators in the corresponding expressions. Note that in Eq. (\ref{11}), the additional power $\omega_{\rm em}$ is due to the one-photon emission probability $W_{2p,1s}^{(1\gamma)}$, as well as different frequency dependence of the level widths in the denominators of Eq. (\ref{12}).
	
	The expression (\ref{12}) can be evaluated using the results of \cite{Ans,Solovyev_2010,Sol-2015}. For a field-quadratic contribution (as a dominant one) to $\Gamma_{\overline{2s}}$, one can write
	\begin{eqnarray}
		\label{13}
		\Gamma_{\overline{2s}} = W_{2p,1s}^{(1\gamma)}\left[\frac{D}{475 {\rm V/cm}}\right]^2.
	\end{eqnarray}
	Then, using the values $W_{2p,1s}^{(1\gamma)} = 6.26826\times 10^8$ s$^{-1}$, $\omega_0= E_{2s}-E_{1s} = 2\,466\,061\,413\,187.035$ kHz, $\Delta E_L = E_{2s}-E_{2p} = 1\,057\,845.0$ kHz \cite{Mohr-2016-RMP} and the field strength $D=10$ V/cm \cite{Parthey}, we find $\delta_a = -2.9$ s$^{-1}$ or $-0.46$ Hz, which is consistent with the estimates above.

	\section{Conclusions}
	
	In this paper, we analyzed the natural asymmetry of the line profile that arises in the process of photon scattering. It is shown that within the framework of the rigorous QED theory, the line shape differs from the "ordinary" Lorentz contour, which is a consequence of the resonance approximation. A more precise description leads to a frequency dependence of the partial transition rates and, as a consequence, of the level widths. Such a deviation, under the assumption that the correction is small, can be regarded as an additional shift in the resonant transition frequency.
	
	In contrast to the known non-resonant corrections and, in particular, the effect of quantum interference, the asymmetry shift occurs even for an "isolated" atomic level. As usual in such studies, the correction due to the natural line profile asymmetry depends on the conditions of experiment. We found that for highly excited states in the hydrogen atom, it does not exceed $10^{-2}$ Hz, i.e. is negligible with respect to the experimental error. However, a rough estimate of the most accurately measured frequency in the hydrogen atom for the $1s-2s$ two-photon transition is close to the experimental error of $10$ Hz \cite{Mat}. Due to the continuous improvement of experiments, it can be expected that such a correction may be of significance in the near future.

	\section*{Acknowledgements}
	The work of A.A. was supported by foundation for the advancement of theoretical physics "BASIS". T. Z. also acknowledges the grant number MK-4796.2022.1.2. 
	\addcontentsline{toc}{section}{List of Literature}
	\bibliographystyle{ieeetr}  
	
	\bibliography{mybibfile}

\begin{thebibliography}{10}

\bibitem{Kennedy}
C.~J. Kennedy, E.~Oelker, J.~M. Robinson, T.~Bothwell, D.~Kedar, W.~R. Milner,
  G.~E. Marti, A.~Derevianko, and J.~Ye, ``Precision metrology meets cosmology:
  Improved constraints on ultralight dark matter from atom-cavity frequency
  comparisons,'' {\em Phys. Rev. Lett.}, vol.~125, p.~201302, Nov 2020.

\bibitem{DIRAC1937}
{Dirac P. A. M.}, ``{The Cosmological Constants},'' {\em Nature}, vol.~139,
  no.~3512, pp.~323--323, 1937.

\bibitem{Webb}
J.~K. Webb, M.~T. Murphy, V.~V. Flambaum, V.~A. Dzuba, J.~D. Barrow, C.~W.
  Churchill, J.~X. Prochaska, and A.~M. Wolfe, ``Further evidence for
  cosmological evolution of the fine structure constant,'' {\em Phys. Rev.
  Lett.}, vol.~87, p.~091301, Aug 2001.

\bibitem{AtCl-Cs}
F.~Levi, D.~Calonico, C.~E. Calosso, A.~Godone, S.~Micalizio, and G.~A.
  Costanzo, ``Accuracy evaluation of {ITCsF}2: a nitrogen cooled caesium
  fountain,'' {\em Metrologia}, vol.~51, pp.~270--284, may 2014.

\bibitem{AtCl-Sr}
T.~L. Nicholson, S.~L. Campbell, R.~B. Hutson, G.~E. Marti, B.~J. Bloom, R.~L.
  McNally, W.~Zhang, M.~D. Barrett, M.~S. Safronova, G.~F. Strouse, W.~L. Tew,
  and J.~Ye, ``Systematic evaluation of an atomic clock at 2 10(-18) total
  uncertainty,'' {\em Nat. Commun.}, vol.~6, p.~6896, Apr 2015.

\bibitem{Parthey}
C.~G. Parthey and et~al., ``Improved measurement of the hydrogen 1s-2s
  transition frequency,'' {\em Phys. Rev. Lett.}, vol.~107, p.~203001, 2011.

\bibitem{Mat}
A.~Matveev, C.~G. Parthey, K.~Predehl, J.~Alnis, A.~Beyer, R.~Holzwarth,
  T.~Udem, T.~Wilken, N.~Kolachevsky, M.~Abgrall, D.~Rovera, C.~Salomon,
  P.~Laurent, G.~Grosche, O.~Terra, T.~Legero, H.~Schnatz, S.~Weyers,
  B.~Altschul, and T.~W. H\"ansch, ``Precision measurement of the hydrogen
  $1s\mathrm{\text{\ensuremath{-}}}2s$ frequency via a 920-km fiber link,''
  {\em Phys. Rev. Lett.}, vol.~110, p.~230801, Jun 2013.

\bibitem{van2011frequency}
R.~Van~Rooij, J.~S. Borbely, J.~Simonet, M.~Hoogerland, K.~Eikema,
  R.~Rozendaal, and W.~Vassen, ``Frequency metrology in quantum degenerate
  helium: Direct measurement of the 2 3s1→ 2 1s0 transition,'' {\em Science},
  vol.~333, no.~6039, pp.~196--198, 2011.

\bibitem{H-exp}
A.~Beyer, L.~Maisenbacher, A.~Matveev, R.~Pohl, K.~Khabarova, A.~Grinin,
  T.~Lamour, D.~C. Yost, T.~W. Hänsch, N.~Kolachevsky, and T.~Udem, ``The
  rydberg constant and proton size from atomic hydrogen,'' {\em Science},
  vol.~358, no.~6359, pp.~79--85, 2017.

\bibitem{Low}
F.~Low, ``Natural line shape,'' {\em Phys. Rev.}, vol.~88, p.~53, 1952.

\bibitem{Andr}
O.~Y. Andreev, L.~N. Labzowsky, G.~Plunien, and D.~A. Solovyev, ``Qed theory of
  the spectral line profile and its applications to atoms and ions,'' {\em
  Phys. Rep.}, vol.~455, pp.~135--246, 2008.

\bibitem{ZSLP-report}
T.~A. Zalialiutdinov, D.~A. Solovyev, L.~N. Labzowsky, and G.~Plunien, ``Qed
  theory of multiphoton transitions in atoms and ions,'' {\em Phys. Rep.},
  vol.~737, pp.~1 -- 84, 2018.

\bibitem{AZSL-2022}
A.~Anikin, T.~Zalialiutdinov, D.~Solovyev, and L.~Labzowsky, ``Line profile
  asymmetry in precision spectroscopy,'' {\em arxiv.2204.12199
  [physics.atom-ph], 26 Aprile 2022}, 2022.

\bibitem{Jent-Mohr}
U.~D. Jentschura and P.~J. Mohr, ``Nonresonant effects in one- and two-photon
  transitions,'' {\em Can. J. of Phys.}, vol.~80, no.~6, pp.~633--644, 2002.

\bibitem{HH-2010}
M.~Horbatsch and E.~A. Hessels, ``Shifts from a distant neighboring
  resonance,'' {\em Phys. Rev. A}, vol.~82, p.~052519, Nov 2010.

\bibitem{HH-2011}
M.~Horbatsch and E.~A. Hessels, ``Shifts from a distant neighboring resonance
  for a four-level atom,'' {\em Phys. Rev. A}, vol.~84, p.~032508, Sep 2011.

\bibitem{Sansonetti}
C.~J. Sansonetti, C.~E. Simien, J.~D. Gillaspy, J.~N. Tan, S.~M. Brewer, R.~C.
  Brown, S.~Wu, and J.~V. Porto, ``Absolute transition frequencies and quantum
  interference in a frequency comb based measurement of the $^{6,7}\mathrm{Li}$
  $d$ lines,'' {\em Phys. Rev. Lett.}, vol.~107, p.~023001, Jul 2011.

\bibitem{Brown-2013}
R.~C. Brown, S.~Wu, J.~V. Porto, C.~J. Sansonetti, C.~E. Simien, S.~M. Brewer,
  J.~N. Tan, and J.~D. Gillaspy, ``Quantum interference and light polarization
  effects in unresolvable atomic lines: Application to a precise measurement of
  the ${}^{6,7}$li ${D}_{2}$ lines,'' {\em Phys. Rev. A}, vol.~87, p.~032504,
  Mar 2013.

\bibitem{MHH-2015}
A.~Marsman, M.~Horbatsch, and E.~A. Hessels, ``The effect of quantum-mechanical
  interference on precise measurements of the n = 2 triplet p fine structure of
  helium,'' {\em J. Phys. Chem. Ref. Data}, vol.~44, no.~3, p.~031207, 2015.

\bibitem{Amaro-2015}
P.~Amaro, F.~Fratini, L.~Safari, A.~Antognini, P.~Indelicato, R.~Pohl, and
  J.~P. Santos, ``Quantum interference shifts in laser spectroscopy with
  elliptical polarization,'' {\em Phys. Rev. A}, vol.~92, p.~062506, Dec 2015.

\bibitem{Amaro-mH-2015}
P.~Amaro, B.~Franke, J.~J. Krauth, M.~Diepold, F.~Fratini, L.~Safari,
  J.~Machado, A.~Antognini, F.~Kottmann, P.~Indelicato, R.~Pohl, and J.~P.
  Santos, ``Quantum interference effects in laser spectroscopy of muonic
  hydrogen, deuterium, and helium-3,'' {\em Phys. Rev. A}, vol.~92, p.~022514,
  Aug 2015.

\bibitem{Anikin}
A.~Anikin, T.~Zalialiutdinov, and D.~Solovyev, ``Angular correlations in
  two-photon spectroscopy of hydrogen,'' {\em Phys. Rev. A}, vol.~103,
  p.~022833, Feb 2021.

\bibitem{Anikin2021}
{Anikin A. A.}, {Zalialiutdinov T. A.}, and {Solovyev D. A.}, ``{Nonresonant
  Effects in the Two-Photon Spectroscopy of a Hydrogen Atom: Application to the
  Calculation of the Charge Radius of the Proton},'' {\em JETP Letters},
  vol.~114, no.~4, pp.~180--187, 2021.

\bibitem{LSPS}
L.~N. Labzowsky, D.~A. Solovyev, G.~Plunien, and G.~Soff, ``Asymmetry of the
  natural line profile for the hydrogen atom,'' {\em Phys. Rev. Lett.},
  vol.~87, p.~143003, Sep 2001.

\bibitem{PRA-LSPS}
L.~Labzowsky, D.~Soloviev, G.~Plunien, and G.~Soff, ``Nonresonant corrections
  to the $1s\ensuremath{-}2s$ two-photon resonance for the hydrogen atom,''
  {\em Phys. Rev. A}, vol.~65, p.~054502, May 2002.

\bibitem{Schwob}
C.~{Schwob}, L.~{Jozefowski}, O.~{Acef}, L.~{Hilico}, B.~{de Beauvoir},
  F.~{Nez}, L.~{Julien}, A.~{Clairon}, and F.~{Biraben}, ``Frequency
  measurement of the 2s-12d transitions in hydrogen and deuterium, new
  determination of the rydberg constant,'' {\em IEEE Transactions on
  Instrumentation and Measurement}, vol.~48, no.~2, pp.~178--181, 1999.

\bibitem{deB-0}
B.~de~Beauvoir, F.~Nez, L.~Julien, B.~Cagnac, F.~Biraben, D.~Touahri,
  L.~Hilico, O.~Acef, A.~Clairon, and J.~J. Zondy, ``Absolute frequency
  measurement of the $2\mathit{S}\ensuremath{-}8\mathit{S}/\mathit{D}$
  transitions in hydrogen and deuterium: New determination of the rydberg
  constant,'' {\em Phys. Rev. Lett.}, vol.~78, pp.~440--443, Jan 1997.

\bibitem{deB-1}
C.~Schwob, L.~Jozefowski, B.~{de Beauvoir}, L.~Hilico, F.~Nez, L.~Julien,
  F.~Biraben, O.~Acef, J.-J. Zondy, and A.~Clairon, ``Optical frequency
  measurement of the 2s-12d transitions in hydrogen and deuterium: Rydberg
  constant and lamb shift determinations,'' {\em Phys. Rev. Lett.}, vol.~82,
  pp.~4960--4963, June 1999.

\bibitem{deB-2}
B.~{de Beauvoir}, C.~Schwob, O.~Acef, L.~Jozefowski, L.~Hilico, F.~Nez,
  L.~Julien, A.~Clairon, and F.~Biraben, ``{Metrology of the hydrogen and
  deuterium atoms: Determination of the Rydberg constant and Lamb shifts},''
  {\em Eur. Phys. J. D}, vol.~12, pp.~61--93, Jan. 2000.

\bibitem{PhysRevA.25.3079}
A.~Quattropani, F.~Bassani, and S.~Carillo, ``Two-photon transitions to excited
  states in atomic hydrogen,'' {\em Phys. Rev. A}, vol.~25, pp.~3079--3089, Jun
  1982.

\bibitem{Eikema}
K.~S.~E. Eikema, J.~Walz, and T.~W. H\"ansch, ``Continuous coherent lyman-
  $\ensuremath{\alpha}$ excitation of atomic hydrogen,'' {\em Phys. Rev.
  Lett.}, vol.~86, pp.~5679--5682, Jun 2001.

\bibitem{LabKlim}
L.~Labzowsky, G.~Klimchitskaya, and Y.~Dmitriev, {\em Relativistic Effects in
  the Spectra of Atomic Systems}.
\newblock Institute of Physics Publishing, 1993.

\bibitem{Zon}
B.~A. {Zon} and L.~P. {Rapoport}, ``{Two-photon Decay of 2s Level of
  Hydrogen},'' {\em ZhETF Pisma Redaktsiiu}, vol.~7, p.~70, Jan. 1968.

\bibitem{Jentschura_2007}
U.~D. Jentschura, ``Non-uniform convergence of two-photon decay rates for
  excited atomic states,'' {\em Journal of Physics A: Mathematical and
  Theoretical}, vol.~40, p.~F223, feb 2007.

\bibitem{GM-1s-3s}
A.~Grinin, A.~Matveev, D.~C. Yost, L.~Maisenbacher, V.~Wirthl, R.~Pohl, T.~W.
  Hänsch, and T.~Udem, ``Two-photon frequency comb spectroscopy of atomic
  hydrogen,'' {\em Science}, vol.~370, no.~6520, pp.~1061--1066, 2020.

\bibitem{Mat-2s-8s}
A.~D. Brandt, S.~F. Cooper, C.~Rasor, Z.~Burkley, A.~Matveev, and D.~C. Yost,
  ``Measurement of the $2{\mathrm{s}}_{1/2}\ensuremath{-}8{\mathrm{d}}_{5/2}$
  transition in hydrogen,'' {\em Phys. Rev. Lett.}, vol.~128, p.~023001, Jan
  2022.

\bibitem{Ans}
Y.~I. Azimov, A.~A. Ansel’m, A.~N. Moskalev, and R.~M. Ryndin, ``Some
  parity-nonconservation effects in emission by hydrogenlike atoms,'' {\em
  JETP}, vol.~40, pp.~8--13, 1975.

\bibitem{Solovyev_2010}
D.~Solovyev, V.~Sharipov, L.~Labzowsky, and G.~Plunien, ``Influence of external
  electric fields on multi-photon transitions between the 2s, 2p and 1s levels
  for hydrogen and antihydrogen atoms and hydrogen-like ions,'' {\em Journal of
  Physics B: Atomic, Molecular and Optical Physics}, vol.~43, p.~074005, mar
  2010.

\bibitem{Sol-2015}
D.~Solovyev and E.~Solovyeva, ``Rydberg-state mixing in the presence of an
  external electric field: Comparison of the hydrogen and antihydrogen
  spectra,'' {\em Phys. Rev. A}, vol.~91, p.~042506, Apr 2015.

\bibitem{Mohr-2016-RMP}
P.~J. Mohr, D.~B. Newell, and B.~N. Taylor, ``Codata recommended values of the
  fundamental physical constants: 2014,'' {\em Rev. Mod. Phys.}, vol.~88,
  p.~035009, Sep 2016.

\end{thebibliography}
	
	\begin{widetext}
		\begin{center}
			\begin{table}
				\caption{Natural asymmetry shift, Eq. (\ref{6}), to the transition frequencies $2s/1s-ns/nd$ ($n=3,\,4,\,6,\,8$) with excited states listed in the first column. In the second and in the third columns the level widths, $\Gamma_{ns/nd}$, and $W^{(1\gamma)}_{ns/nd\rightarrow kp}$ are listed, respectively. In the fourth and in the fifth columns resonant transition frequencies and the energy intervals $\Delta E_{ns/nd,kp}$ are listed, respectively. Finally, in the last two columns the first and the second terms of $\delta_a$ (see Eq. (\ref{6})) are given. The total values of $\delta_a$ are listed in the substring with empty cells. All values are obtained in a purely nonrelativistic theory and are given in Hz.}
				\label{tab:1}
				\begin{tabular}{l | c | c | c | c | c | r}
					\hline
					\hline
					state & $\Gamma_{ns/nd}$ & $W^{(1\gamma)}_{ns/nd\rightarrow kp}$ & $\omega_0$ & $\Delta E_{ns/nd,kp}$ & $\frac{3}{4}\frac{\Gamma_{ns/nd}^2}{\omega_0}$ & $-\frac{\Gamma_{ns/nd}}{4}\frac{W^{(1\gamma)}_{ns/nd\rightarrow kp}}{\Delta E_{ns/nd,kp}}$\\
					\hline
					
					$3s$ & $1.005\times 10^6$ & $1.005\times 10^6$ (2p) & $2.924\times 10^{15}$ & $4.569\times 10^{14}$ & $2.59\times 10^{-4}$ & $-5.53\times 10^{-4}$ \\
					& & & & & & $-2.94\times 10^{-4}$ \\
					\hline
					
					$3d$ & $1.029\times 10^7$ & $1.029\times 10^7$ (2p) & $2.924\times 10^{15}$ & $4.569\times 10^{14}$ & $2.72\times 10^{-2}$ & $-5.79\times 10^{-2}$ \\
					& & & & & & $-3.08\times 10^{-2}$ \\
					\hline

					$4s$ & $7.028\times 10^5$ & $2.923\times 10^5$ (3p) & $6.168\times 10^{14}$ & $1.599\times 10^{14}$ & $6.01\times 10^{-4}$ & $-3.21\times 10^{-4}$ \\
					& $7.028\times 10^5$ & $4.105\times 10^5$ (2p) & $6.168\times 10^{14}$ & $6.168\times 10^{14}$ & $6.01\times 10^{-4}$ & $-1.17\times 10^{-4}$ \\
					
					& & & & & & $1.62\times 10^{-4}$ \\
					\hline
					
					$4d$ & $4.405\times 10^6$ & $1.121\times 10^6$ (3p) & $6.168\times 10^{14}$ & $1.599\times 10^{14}$ & $2.36\times 10^{-2}$ & $-7.72\times 10^{-3}$ \\
					& $4.405\times 10^6$ & $3.284\times 10^6$ (2p) & $6.168\times 10^{14}$ & $6.168\times 10^{14}$ & $2.36\times 10^{-2}$ & $-5.86\times 10^{-3}$ \\
					
					& & & & & & $1.00\times 10^{-2}$ \\
					\hline
					
					$6s$ & $2.975\times 10^5$ & $4.270\times 10^4$ (5p) & $7.311\times 10^{14}$ & $4.021\times 10^{13}$ & $9.08\times 10^{-5}$ & $-7.90\times 10^{-5}$ \\
					& $2.975\times 10^5$ & $5.705\times 10^4$ (4p) & $7.311\times 10^{14}$ & $1.142\times 10^{14}$ & $9.08\times 10^{-5}$ & $-3.72\times 10^{-5}$ \\
					& $2.975\times 10^5$ & $8.076\times 10^4$ (3p) & $7.311\times 10^{14}$ & $2.741\times 10^{14}$ & $9.08\times 10^{-5}$ & $-2.19\times 10^{-5}$ \\
					& $2.975\times 10^5$ & $1.170\times 10^5$ (2p) & $7.311\times 10^{14}$ & $7.311\times 10^{14}$ & $9.08\times 10^{-5}$ & $-1.19\times 10^{-5}$ \\     
					& & & & & & $-5.92\times 10^{-5}$ \\
					\hline
					
					$6d$ & $1.337\times 10^6$ & $7.158\times 10^4$ (5p) & $7.311\times 10^{14}$ & $4.021\times 10^{13}$ & $1.83\times 10^{-3}$ & $-5.95\times 10^{-4}$ \\
					& $1.337\times 10^6$ & $1.373\times 10^5$ (4p) & $7.311\times 10^{14}$ & $1.142\times 10^{14}$ & $1.83\times 10^{-3}$ & $-4.02\times 10^{-4}$ \\
					& $1.337\times 10^6$ & $2.990\times 10^5$ (3p) & $7.311\times 10^{14}$ & $2.741\times 10^{14}$ & $1.83\times 10^{-3}$ & $-3.65\times 10^{-4}$ \\
					& $1.337\times 10^6$ & $8.193\times 10^5$ (2p) & $7.311\times 10^{14}$ & $7.311\times 10^{14}$ & $1.83\times 10^{-3}$ & $-3.75\times 10^{-4}$ \\     
					& & & & & & $9.77\times 10^{-5}$ \\
					\hline
					
					$8s$ & $1.441\times 10^5$ & $1.046\times 10^4$ (7p) & $7.710\times 10^{14}$ & $1.573\times 10^{13}$ & $2.02\times 10^{-5}$ & $-2.40\times 10^{-5}$ \\
					& $1.441\times 10^5$ & $1.290\times 10^4$ (6p) & $7.710\times 10^{14}$ & $3.998\times 10^{13}$ & $2.02\times 10^{-5}$ & $-1.16\times 10^{-5}$ \\
					& $1.441\times 10^5$ & $1.656\times 10^4$ (5p) & $7.710\times 10^{14}$ & $8.019\times 10^{13}$ & $2.02\times 10^{-5}$ & $-7.44\times 10^{-6}$ \\
					& $1.441\times 10^5$ & $2.260\times 10^4$ (4p) & $7.710\times 10^{14}$ & $1.542\times 10^{14}$ & $2.02\times 10^{-5}$ & $-5.28\times 10^{-6}$ \\   
					& $1.441\times 10^5$ & $3.291\times 10^4$ (3p) & $7.710\times 10^{14}$ & $3.141\times 10^{14}$ & $2.02\times 10^{-5}$ & $-3.77\times 10^{-6}$ \\
					& $1.441\times 10^5$ & $4.863\times 10^4$ (2p) & $7.710\times 10^{14}$ & $7.710\times 10^{14}$ & $2.02\times 10^{-5}$ & $-2.27\times 10^{-6}$ \\   
					& & & & & & $-3.41\times 10^{-5}$ \\
					\hline
					
					$8d$ & $5.724\times 10^5$ & $1.201\times 10^4$ (7p) & $7.710\times 10^{14}$ & $1.573\times 10^{13}$ & $3.19\times 10^{-5}$ & $-1.09\times 10^{-4}$ \\
					& $5.724\times 10^5$ & $1.878\times 10^4$ (6p) & $7.710\times 10^{14}$ & $3.998\times 10^{13}$ & $3.19\times 10^{-5}$ & $-6.72\times 10^{-5}$ \\
					& $5.724\times 10^5$ & $3.079\times 10^4$ (5p) & $7.710\times 10^{14}$ & $8.019\times 10^{13}$ & $3.19\times 10^{-5}$ & $-5.49\times 10^{-5}$ \\
					& $5.724\times 10^5$ & $5.612\times 10^4$ (4p) & $7.710\times 10^{14}$ & $1.542\times 10^{14}$ & $3.19\times 10^{-5}$ & $-5.21\times 10^{-5}$ \\   
					& $5.724\times 10^5$ & $1.204\times 10^5$ (3p) & $7.710\times 10^{14}$ & $3.141\times 10^{14}$ & $3.19\times 10^{-5}$ & $-5.49\times 10^{-5}$ \\
					& $5.724\times 10^5$ & $3.268\times 10^5$ (2p) & $7.710\times 10^{14}$ & $7.710\times 10^{14}$ & $3.19\times 10^{-5}$ & $-6.07\times 10^{-5}$ \\   
					& & & & & & $-8.03\times 10^{-5}$ \\
					\hline
					\hline
				\end{tabular}
				\label{tab1}
			\end{table}
		\end{center}
	\end{widetext}

\end{document}